\documentclass[12pt]{iopart}

\usepackage{epsfig}
\usepackage{rotating}

\begin{document}

\title[On the properties of surface reconstructed SiNWs]{On the properties
       of surface reconstructed silicon nanowires} 

\author{R Rurali\footnote[3]{To whom correspondence should be addressed 
                             (rurali@irsamc.ups-tlse.fr)}
        and N Lorente}

\address{Laboratoire Collisions, Agr\'{e}gats, R\'{e}activit\'{e},
         IRSAMC, Universit\'{e} Paul Sabatier, \\
         118 route de Narbonne, 31062 Toulouse cedex,
         France} 

\begin{abstract}

We study by means of density-functional calculations the role of lateral
surface reconstructions in determining the electrical properties
of $\langle 100 \rangle$ silicon nanowires. The different lateral 
reconstructions are explored by relaxing all the nanowires with 
crystalline bulk silicon structure and all possible ideal facets that 
correspond to an average diameter of 1.5~nm. We show that the 
reconstruction induces the formation of ubiquitous surface states
that turn the wires into semi-metallic or metallic.

\end{abstract}




\section{Introduction}
\label{sec:intro}

There has been a growing interest in semiconductor 
nanowires~\cite{appell,morales,lieber}
for their potential use in future nanoelectronic applications, 
such as nanocontacts and nanoswitches. Silicon nanowires~(SiNWs) are 
especially attractive for their possible efficient integration in 
conventional Si-based microelectronics. 
The use of SiNWs as chemical sensors has also been demonstrated.
SiNW-based sensors for the detection of NH$_3$~\cite{zhou}, of biological 
macromolecules~\cite{cui} and for the identification of complementary 
vs. mismatched DNA~\cite{hahm} have been reported.
Experimentally, SiNWs are grown from a nanocrystal~\cite{morales,holmes} 
which is used as a seed to direct the one-dimensional crystallisation 
of silicon. The (typically gold) nanocluster serves as the critical point 
for nucleation and enables the addition of reactants that allow the growth, 
determining the nanowire diameter and orientation. Ma {\em et al.}~\cite{ma} 
achieved the thinnest SiNWs reported so far, with diameters as small as 
1.3~nm for a wire grown along the $\langle 110 \rangle$ direction. 
A few other groups~\cite{holmes,wu,colemann} also obtained SiNWs with 
diameters below 10~nm and growth orientations including  
$\langle 100 \rangle$~\cite{holmes}, $\langle 110 \rangle$~\cite{holmes,wu}, 
$\langle 111 \rangle$~\cite{wu} and $\langle 112 \rangle$~\cite{wu}.

The one-dimensionality is known to induce a gap broadening effect,
due to quantum confinement, in H-passivated SiNWs~\cite{delley}.
In this paper we explore on theoretical grounds the electronic structure 
of SiNWs when the lateral surface is left free to reconstruct. 
Silicon surfaces are among the most studied systems of the latest decades, 
both theoretically and experimentally, and it is well-known that the
reconstruction drastically affects their electronic properties.
In Section~\ref{sub:wulff} we discuss the geometry of the surface 
reconstructed SiNWs considered and the implication of the Wulff's rule
in the case nanoscale one-dimensional systems.  
In Section~\ref{sub:electronic-structure} we analyse the electronic
structure of the most favoured reconstruction and the localisation
of the surface states that have been found to form.

\section{Computational methods}
\label{sec:computational}

The calculations presented in this paper have been carried out in 
the framework of density-functional theory~(DFT). We have used both
a numerical atomic orbital~\cite{siesta} and a plane-wave~\cite{dacapo}
basis set. We have used a double-$\zeta$ polarised basis set~\cite{siesta}
with pseudopotentials of the Troullier-Martins~\cite{troullier:martins} 
type and a plane-wave energy cutoff of 20~Ry~\cite{dacapo} with ultrasoft 
pseudopotentials~\cite{vanderbilt}.
In both cases the exchange-correlation energy was calculated according to 
the Generalised Gradient Approximation~\cite{gga}.
The wires that we have studied have a diameter of $\sim 1.5$~nm and a number
of atoms ranging from 57 to 171, depending on the supercell size and
on the adopted shape of the unrelaxed wire section.
The Brillouin zone has been sampled according to the 
Monkhorst-Pack~\cite{monkhorst:pack} scheme with a converged grid of
$1 \times 1 \times 4$, $1 \times 1 \times 6$ or $1 \times 1 \times 12$ 
k-points, depending on the supercell size.

\begin{figure}
\begin{center}
\epsfxsize=5cm
\epsffile{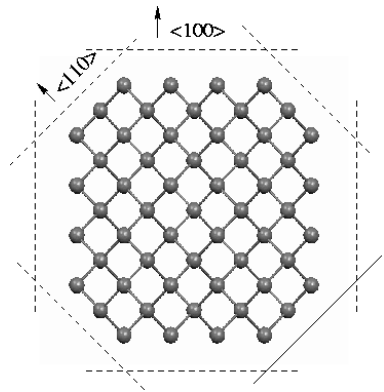}
\end{center}
\caption{Facet arrangement of a $\langle 100 \rangle$ SiNW as dictated
         by the Wulff's rule. The formation of \{110\} facets allows a
         smoother matching between \{100\} facets.}
\label{fig:wulff}
\end{figure}

\section{Results and discussion}
\label{sec:results}

\subsection{Structural relaxations: checking the Wulff's rule at the nanoscale}
\label{sub:wulff}

\begin{figure}
\begin{center}
\epsfxsize=8cm
\epsffile{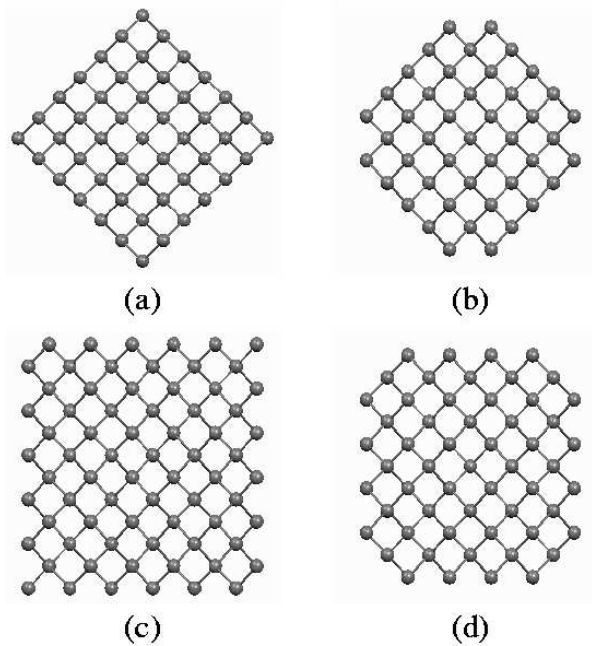}
\end{center}
\caption{Unrelaxed section of the SiNWs considered.}
\label{fig:unrelax}
\end{figure}

Convincing experimental evidence shows that SiNWs are constructed 
around a crystalline bulk core~\cite{ma,zhang}. The shape of the 
section of such a structure will depend on the way in which the involved 
vicinal surfaces match and on how abrupt is the transition from one 
to the other. The case of SiNWs grown along the $\langle 100 \rangle$ 
direction is of especial interest, because it determines a rather 
conflictive situation at the edges where two (100) surfaces meet with 
a $90^\circ$ angle [see Figure~\ref{fig:wulff} and \ref{fig:unrelax}(c)]. 
In such conditions, it is known from Wulff's theorem~\cite{wulff} that the 
formation of a {\em facet} is to be 
expected, because it would partly release the stress accumulated 
at the edge. An example of {\em faceting} of a $\langle 100 \rangle$ 
SiNW~\cite{arias,100} is shown in Figure~\ref{fig:wulff}. 
As can be seen, the formation of a 
\{110\} facet allows a much smoother transition between the two 
vicinal (100)-like surfaces. However, as discussed previously by Zhang and 
Yakobson~\cite{yakobson}, in the case of very thin one-dimensional 
structures such as those that we are treating, the predictions of Wulff's rule 
should be carefully revised. The size of the \{110\} and of the 
\{100\} facets are of the same order of magnitude, thus the surface energy 
associated to the edges can no longer be neglected, as assumed in Wulff's rule.
For these reasons, the first part of this work has consisted in
identifying the favoured faceting arrangement of $\langle 100 \rangle$ 
SiNW of nanometric thickness, thus verifying the validity of Wulff's 
theorem in these conditions. 

We have considered two different square-section wires 
[Figure~\ref{fig:unrelax}(a) and (c)] and two Wulff faceting 
arrangements [Figure~\ref{fig:unrelax}(b) and (d)]. 
The square wire of Figure~\ref{fig:unrelax}(a) is made of \{110\} facets, 
while the wire in Figure~\ref{fig:unrelax}(c) has only \{100\} facets. 
The wires of Figure~\ref{fig:unrelax}(b) and (d) have been obtained from 
those of Figure~\ref{fig:unrelax}(a) and (c), respectively, smoothening the 
corners according to Wulff's rule prescription. The way in which the 
tension at the edges is released and the way in which the different 
Si surface reconstructions (\{100\} and \{110\} facets) compete among 
them will determine the minimum energy structure. Therefore, given the 
starting configurations of Figure~\ref{fig:unrelax}, very different 
minimum energy structures are expected.

\begin{figure}
\begin{center}
\epsfxsize=8cm
\epsffile{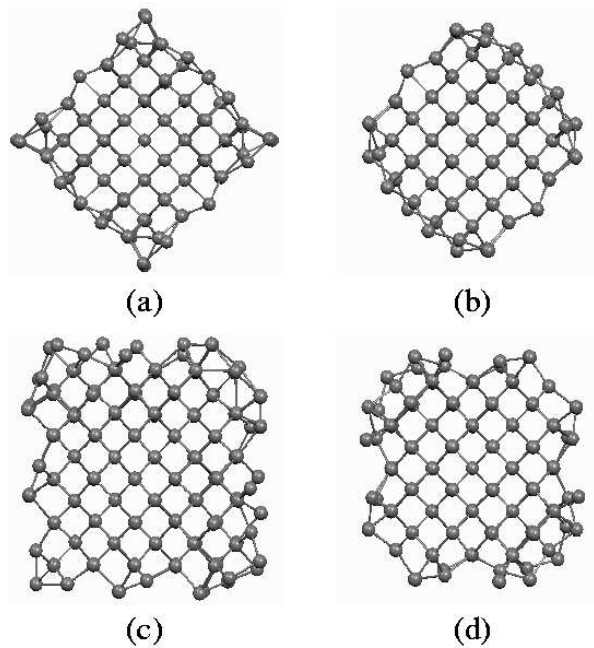}
\end{center}
\caption{Relaxed section of the SiNWs considered.}
\label{fig:relax}
\end{figure}

The results of the relaxations are shown in the section-view of 
Figure~\ref{fig:relax}. As can be seen, all but one of the wires maintain 
a high degree of in-plane symmetry. As expected, in the SiNW of 
Figure~\ref{fig:relax}(a) \{110\} facets are dominant. 
The same relaxation pattern 
is followed by the wire of Figure~\ref{fig:relax}(b), with the only notable 
exception of the missing corner atoms, whose removal allow a smoother match 
between the vicinal (110)-like surfaces. In accordance to the prediction 
of Wulff's rule, the SiNW of Figure~\ref{fig:relax}(b) turned out to be more 
stable than that of Figure~\ref{fig:unrelax}(a) (the difference amounting 
to $\sim$~33~meV/atom).

In the relaxation of the SiNWs of Figure~\ref{fig:relax}(c) and (d) 
\{100\} facets prevail. It is interesting to note that in this case 
the effects of a Wulff-like profile are more evident than in the former case. 
Here an actual edge-faceting develops, mediating the transition between
the vicinal (100)-like surfaces by \{110\} facets 
[see Figure~\ref{fig:relax}(d)].
On the other hand, when the edges between the \{100\} facets are not 
smoothened, the wire favours a symmetry breaking, ending up with the 
{\em butterfly}-shaped section of Figure~\ref{fig:relax}(c). 
Interestingly enough, the {\em butterfly} wire of Figure~\ref{fig:relax}(c)
has turned out to be slightly more stable than the wire of
Figure~\ref{fig:relax}(d), thus confirming that at such small diameters
the validity of Wulff's rule should be carefully checked.

Regarding the influence of facet orientation, we have found that wires 
dominated by \{100\} facets [Figure~\ref{fig:relax}(c) and (d)] are more stable 
than those where \{110\} facets prevail [(Figure~\ref{fig:relax}(a) and (b)]. 
The cohesive energies are summarised in Table~\ref{tab:cohesive}.
For these reasons, the candidate structures for SiNWs grown along the 
\{100\} direction appear to be those in Figure~\ref{fig:relax}(c) and (d),
favouring the formation of \{100\} over \{110\} facets.

\begin{table}
\begin{center}
\begin{tabular}{ccc}
\hline\hline
Dominant facets & Section shape & Cohesive energy (eV/atom) \\
\hline
$\langle 110 \rangle$ & square & 3.886 \\
$\langle 110 \rangle$ & Wulff  & 3.919 \\
$\langle 100 \rangle$ & square & 4.006 \\
$\langle 100 \rangle$ & Wulff  & 3.989 \\
\end{tabular}
\end{center}
\caption{Cohesive energies per atoms of the SiNW type illustrated in 
         Fig.~\ref{fig:relax}. The SiNWs where \{100\} facets prevail
         are approximately 0.1~eV/atom more stable.}
\label{tab:cohesive}
\end{table}

\begin{figure}
\begin{center}
\epsfxsize=7cm
\epsffile{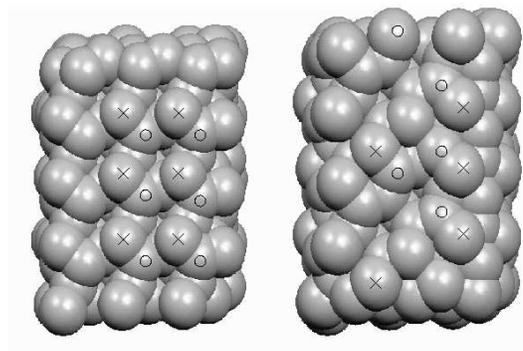}
\end{center}
\caption{Minimum energy geometries for the facets of $\langle 100 \rangle$ 
         {\em butterfly}-like SiNW of Figure~\ref{fig:relax}(d).} 
\label{fig:facets}
\end{figure}

We have found that for the SiNW of Figure~\ref{fig:relax}(d) two different
reconstructions of the \{100\} facets are possible~\cite{100}.
Like in the case of the infinite Si(100) surface, the reconstruction 
consists in the formation of a sequence of buckling dimers. What differs
between the two competing geometries that we have obtained is the 
pattern followed by the Si dimers. In one case they determine a {\em trough} 
in the middle of the facet, while in the other one every two dimers on one 
of the two sides is {\em flipped} with respect to the symmetric behaviour 
of the {\em trough} reconstruction.
The relaxed geometry of the facets of the {\em butterfly} wire is shown
in Figure~\ref{fig:facets}. The symmetry breaking is reflected in the 
relaxation and the \{100\} facets follows two different relaxation patterns. 
One of them -~Figure~\ref{fig:facets}(a)~- is rather symmetric, with all
the dimers of one side flipped; the facet of Figure~\ref{fig:facets}(b) 
presents a regular sequence of buckled dimers only in one of the two sides, 
while in the other one every two dimers is missing.

\subsection{Electronic structure}
\label{sub:electronic-structure}

Nanowires -~and SiNWs in particular~- are expected to play an important
role in future molecular electronics applications. Therefore, a thorough
understanding of their conductive properties is required. In this section
we discuss our results of the electronic structure of the {\em butterfly} 
wire [Figure~\ref{fig:relax}(c)], as well as of the two different
reconstructions that we have found for the Wulff-like SiNW 
[Figure~\ref{fig:relax}(d)].

\begin{figure}
\begin{center}
\epsfxsize=7cm
\epsffile{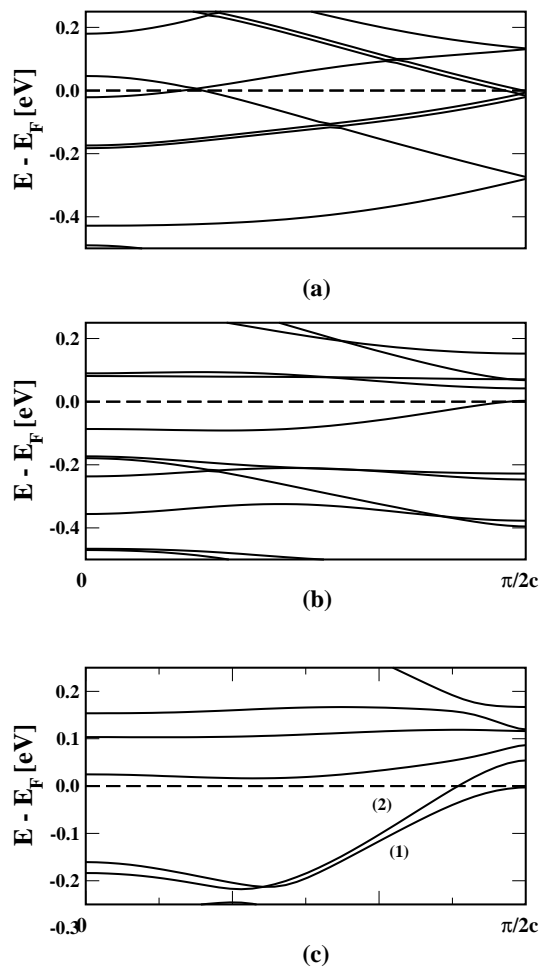}
\end{center}
\caption{Band structure diagrams of (a) the {\em trough} and (b) the 
         {\em dimer flipped} reconstruction of the SiNW with a Wulff-like
         section; (c) the {\em butterfly}-like SiNW.}
\label{fig:bands}
\end{figure}

In Figure~\ref{fig:bands}(a) and (b) the band structure diagrams 
corresponding to 
the two competing geometries for the Wulff-like wire are displayed. It 
is somehow surprising to discover that they are rather different: while
the {\em trough} reconstruction is strongly metallic, with four bands
crossing the Fermi level, the {\em flipped dimer} geometry is only
semi-metallic, with one band approaching the Fermi energy with a zero derivative
at the zone boundary. This is quite striking, at first sight, because
the differences between the two reconstructions do not seem to be so 
relevant. However, when considered in more detail, one notices the pivotal
role that the flipped dimer has in breaking the surface Bloch state that 
would otherwise form. On the contrary, the surface state forms in the 
trough reconstruction where all the Si dimers follow the same buckling 
pattern.

\begin{figure}
\begin{center}
\epsfxsize=12cm
\epsffile{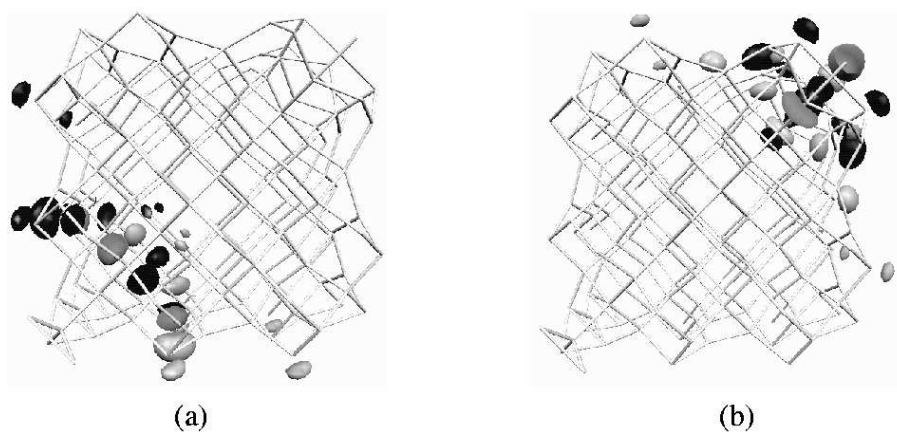}
\end{center}
\caption{Wave function of (a)~the semi-metallic and (b)~metallic state, 
         respectively labelled as (1) and (2) in Figure~\ref{fig:bands}(c).}
\label{fig:wf-butter}
\end{figure}

The band structure of the {\em butterfly} wire is shown in 
Figure~\ref{fig:bands}(c). Also in this case the reconstruction results
in a metallisation of the wire surface, with one band crossing the Fermi
level and a semi-metallic band tangent to it at the zone boundary.
The formation of these metallic and semi-metallic states is directly 
induced by the reconstruction of the surface dimers. Therefore, they
are expected to be localised at the wire's outer layers. The dimerisation 
leads to a dangling bond that hybridises with the other dimers' dangling 
bonds along the nanowire in a surface $\pi$-bond. The surface nature
of the semi-metallic and metallic states of Figure~\ref{fig:bands}(c)
is evident in Figure~\ref{fig:wf-butter} where we have plotted the 
corresponding wave functions. As can be seen, the metallic and semi-metallic 
state are localised, at opposite sides, along one of the $\langle 110 \rangle$ 
diagonal. A similar surface localisation, though more symmetric, following
the pattern of the overall relaxed geometry, is observed for 
the Wulff-like SiNW of
Figure~\ref{fig:relax}(d) (not shown here; see Reference~\cite{100}).
Therefore, conduction in surface-reconstructed SiNWs will be almost
entirely sustained by the outer layers of the wire, with a negligible
penetration inside the wire's core. These results, i.e.~metallic or 
semi-metallic nature and surface character of the conduction channels,
concern all of the $\langle 100 \rangle$ wire types that are likely
to be obtained. As we have discussed in Section~\ref{sub:wulff}
(see also Table~\ref{tab:cohesive}), the difference of cohesive
energies among the favoured wire geometries [the wire of 
Figure~\ref{fig:relax}(c) and the two reconstructions of 
Figure~\ref{fig:relax}(d)] are very small and the selective growth 
of one or another wire seems a difficult task. 

\section{Conclusions}

We have shown that in absence of a proper passivation, the lateral
surface of SiNWs strongly reconstructs, forming series of buckled
dimers with a pattern similar to Si (100) surfaces. 
Depending on the reconstruction, surface states that cross the Fermi level
can form, leading to metallic or semi-metallic SiNWs. Under such circumstances,
doping is no longer needed to have highly conducting nanowires.
The possibility of tailoring SiNWs that are conducting without the need 
of doping is particularly relevant in very thin wires where typical doping 
concentrations require a precision in the fraction of impurities per atom
extremely difficult to control.

Available experiments have not thoroughly explored
the possibility of surface reconstruction or nanowire conductance
without doping. However, these results show a promising venue of
experimental and theoretical research of pure SiNWs.

\vspace{0.25in}

R.R. acknowledges the financial support of the Generalitat de Catalunya
through a {\sc Nanotec} fellowship and thanks N.~Bedoya for the help in 
running some calculations. N.L. thanks ACI jeunes chercheurs.
Computational resources at the Centre Informatique National de l'Enseignement
Sup\'erieur and the Centre de Calcul Midi-Pyr\'en\'ees are gratefully
acknowledged.

\end{document}